\documentclass[twocolumn, switch]{article} % Method A for two-column formatting

\usepackage{preprint}

\usepackage{amsmath, amsthm, amssymb, amsfonts}

\usepackage[numbers,square]{natbib}
\usepackage[utf8]{inputenc}	% allow utf-8 input
\usepackage[T1]{fontenc}	% use 8-bit T1 fonts
\usepackage{xcolor}		% colors for hyperlinks
\usepackage[colorlinks = true,
            linkcolor = purple,
            urlcolor  = blue,
            citecolor = cyan,
            anchorcolor = black]{hyperref}	% Color links to references, figures, etc.
\usepackage{booktabs} 		% professional-quality tables
\usepackage{nicefrac}		% compact symbols for 1/2, etc.
\usepackage{microtype}		% microtypography
\usepackage{lineno}		% Line numbers
\usepackage{float}			% Allows for figures within multicol

\usepackage{lipsum}		%  Filler text

\usepackage{newfloat}
\DeclareFloatingEnvironment[name={Supplementary Figure}]{suppfigure}
\usepackage{sidecap}
\sidecaptionvpos{figure}{c}

\usepackage{titlesec}
\titlespacing\section{0pt}{12pt plus 3pt minus 3pt}{1pt plus 1pt minus 1pt}
\titlespacing\subsection{0pt}{10pt plus 3pt minus 3pt}{1pt plus 1pt minus 1pt}
\titlespacing\subsubsection{0pt}{8pt plus 3pt minus 3pt}{1pt plus 1pt minus 1pt}

\usepackage{amsmath,amsfonts}
\usepackage{hyperref}
\usepackage{wrapfig}
\usepackage{color,colortbl}
\usepackage{tabu}
\usepackage{arydshln}
\usepackage{diagbox}
\usepackage{pifont}
\usepackage{makecell}% to use \Xhline{1pt}
\usepackage{multirow}
\usepackage{xcolor}
\usepackage{bm}
\usepackage[capitalize]{cleveref}
\hypersetup{hidelinks}
\usepackage{xcolor}
\usepackage{overpic}
\usepackage{subfigure}
\crefformat{section}{\S#2#1#3}
\crefformat{subsection}{\S#2#1#3}
\crefformat{subsubsection}{\S#2#1#3}
\crefformat{table}{Table~#2#1#3}

\definecolor{myred}{RGB}{255,0,0}
\definecolor{mygreen}{RGB}{0,255,0}
\definecolor{myblue}{RGB}{0,0,255}

\newcommand{\etc}{\textsl{etc}}

\newcommand{\ie}{\textit{i}.\textit{e}.}

\usepackage{setspace}

\usepackage{tikz,xcolor,hyperref}

\definecolor{lime}{HTML}{A6CE39}
\DeclareRobustCommand{\orcidicon}{
	\begin{tikzpicture}
	\draw[lime, fill=lime] (0,0)
	circle [radius=0.16]
	node[white] {{\fontfamily{qag}\selectfont \tiny ID}};
	\draw[white, fill=white] (-0.0625,0.095)
	circle [radius=0.007];
	\end{tikzpicture}
	\hspace{-2mm}
}
\foreach \x in {A, ..., Z}{\expandafter\xdef\csname orcid\x\endcsname{\noexpand\href{https://orcid.org/\csname orcidauthor\x\endcsname}
			{\noexpand\orcidicon}}
}
\title{MARformer: An Efficient Metal Artifact Reduction Transformer for Dental CBCT Images}

\usepackage{authblk}

\author[1]{Yuxuan Shi}
\author[1\thanks{\tt{csjunxu@nankai.edu.cn}}]{Jun Xu}
\author[2]{Dinggang Shen}

\affil[1]{School of Statistics and Data Science, Nankai University}
\affil[2]{School of Biomedical Engineering, ShanghaiTech University}

\begin{document}

\twocolumn[ % Method A for two-column formatting
  \begin{@twocolumnfalse} % Method A for two-column formatting

\maketitle

\begin{abstract}
Cone Beam Computed Tomography (CBCT) plays a key role in dental diagnosis and surgery. However, the metal teeth implants could bring annoying metal artifacts during the CBCT imaging process, interfering diagnosis and downstream processing such as tooth segmentation.
In this paper, we develop an efficient Transformer to perform metal artifacts reduction (MAR) from dental CBCT images. The proposed MAR Transformer (MARformer) reduces computation complexity in the multihead self-attention by a new Dimension-Reduced Self-Attention (DRSA) module, based on that the CBCT images have globally similar structure.
A Patch-wise Perceptive Feed Forward Network (P2FFN) is also proposed to perceive local image information for fine-grained restoration. Experimental results on CBCT images with synthetic and real-world metal artifacts show that our MARformer is efficient and outperforms previous MAR methods and two restoration Transformers. 
\end{abstract}
% \keywords{Metal Artifact Reduction \and Transformer \and Dimension Reduction} % (optional)
\vspace{0.35cm}

  \end{@twocolumnfalse} % Method A for two-column formatting
] % Method A for two-column formatting

%\begin{multicols}{2} % Method B for two-column formatting (doesn't play well with line numbers), comment out if using method A

%%%%%%%%%%%%%%%  Main text   %%%%%%%%%%%%%%%
% \linenumbers
\section{Introduction}
\begin{figure}[t]
	\centering
	\begin{minipage}[t]{0.23\textwidth}
		\centering 
		\includegraphics[width=0.95\textwidth]{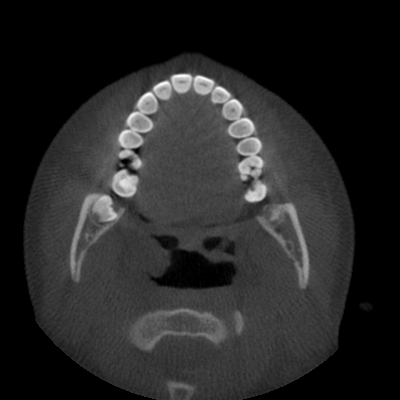}
		{\scriptsize (a) MA-free Image}
	\end{minipage}
	\hfill
	\begin{minipage}[t]{0.23\textwidth}
		\centering 
		\includegraphics[width=0.95\textwidth]{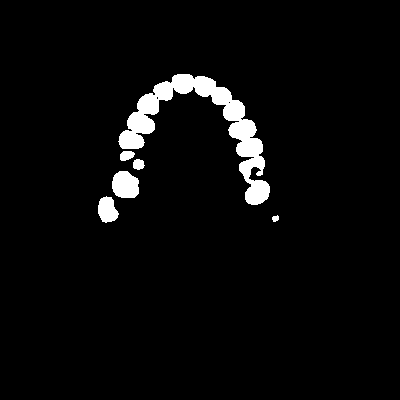}
		{\scriptsize (b) Segmentation on (a)}
	\end{minipage}
	\hfill
	\begin{minipage}[t]{0.23\textwidth}
		\centering 
		\includegraphics[width=0.95\textwidth]{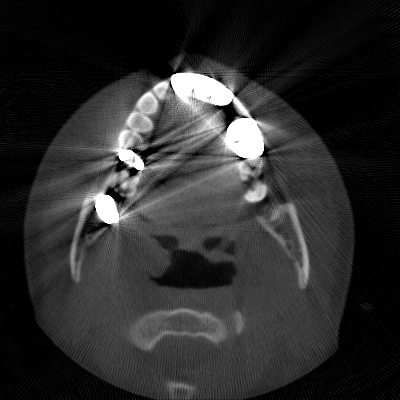}
		{\scriptsize (c) MA Image}
	\end{minipage}
	\hfill
	\begin{minipage}[t]{0.23\textwidth}
		\centering 
		\includegraphics[width=0.95\textwidth]{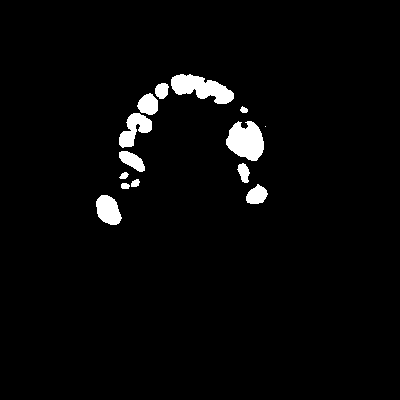}
		{\scriptsize (d) Segmentation on (c)}
	\end{minipage}
	\caption{Segmentation results by Poolformer~\citep{yu2022metaformer}. (a) A CBCT image without metal artifacts. (b) Segmentation mask of Poolformer on (a). (c) The image (a) with synthetic metal artifacts. (d) Segmentation mask of Poolformer on (c).}
	\label{segmentation-example}
\end{figure}

%As
As a useful technology for radiographic imaging, Cone Beam Computed Tomography (CBCT) is widely utilized in oral surgery~\citep{weiss2019cone}, 
orthodontics~\citep{kapila2015cbct}, 
and dental implant~\citep{song2009correlation}, \etc. High-quality dental CBCT imaging is of great importance to guarantee accurate clinical diagnosis. However, the metal teeth implants would bring unpleasant metal artifacts like flare during the imaging process due to the radiation scattering and beam hardening effects~\citep{mouton2013experimental}. 
These artifacts will somewhat damage the post-processing diagnosis-related tasks like tooth segmentation. For example, in \cref{segmentation-example}, we present two CBCT images with or without flare artifacts contaminated by metals in teeth. 
The two images are segmented by a segmentation network~\citep{yu2022metaformer} trained on large scale tooth CBCT images with accurate masks annotated by experienced dentists. One can see that, the CBCT image with metal artifacts are inaccurately segmented with heavy missing of teeth area, which cannot be used to infer the teeth structure. To this end, it is essential to remove the metal artifacts on the dental CBCT images with metal implants for the following usage of tooth segmentation in clinical diagnosis.

\begin{figure*}[ht]
    \centering
    \begin{subfigure}
        \centering
        \begin{minipage}[t]{\textwidth}
            \centering
            \begin{overpic}[width=\textwidth]{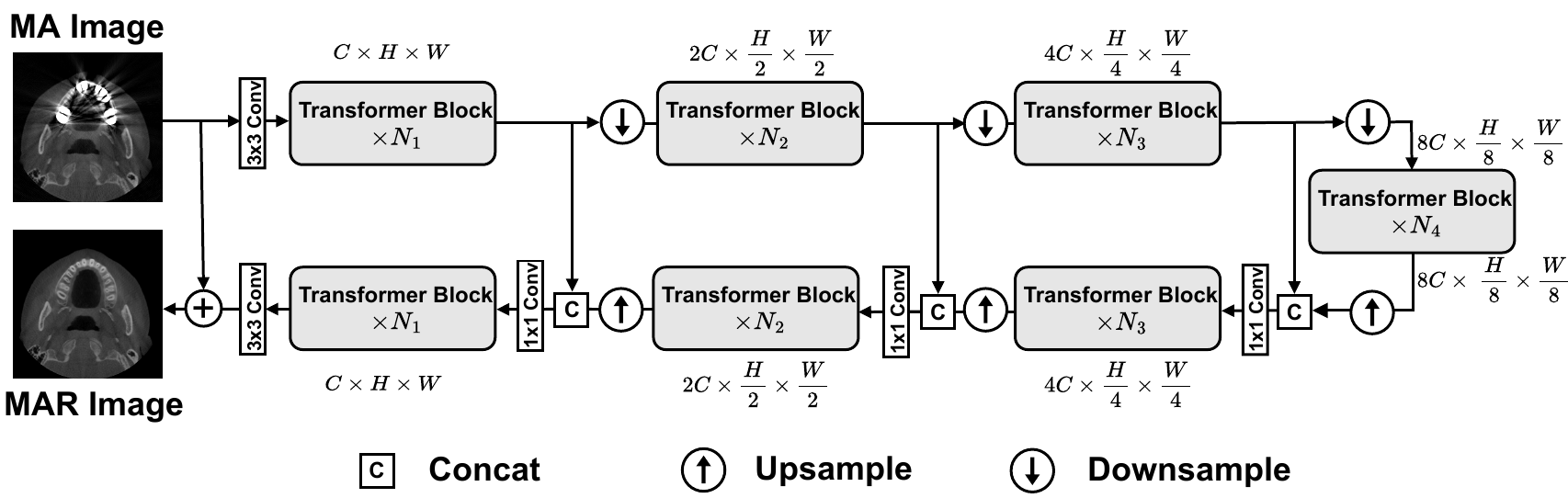}
            \put(6.5,1){\scriptsize (a)}
            \end{overpic}
            % \raisebox{-2cm}{\includegraphics[width= \textwidth]{images/MARformer.pdf} }
            % {\scriptsize (a)}
        \end{minipage}

    \end{subfigure}
    
    \begin{subfigure}
        \centering
        \begin{minipage}[t]{0.14 \textwidth}
            \centering
            \raisebox{-2cm}{\includegraphics[width=1 \textwidth]{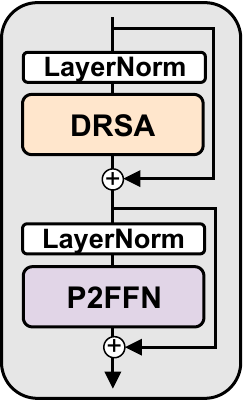} }
            {\scriptsize (b)}
        \end{minipage}
        \begin{minipage}[t]{0.62 \textwidth}
            \centering
            \raisebox{-2cm}{\includegraphics[width=1 \textwidth]{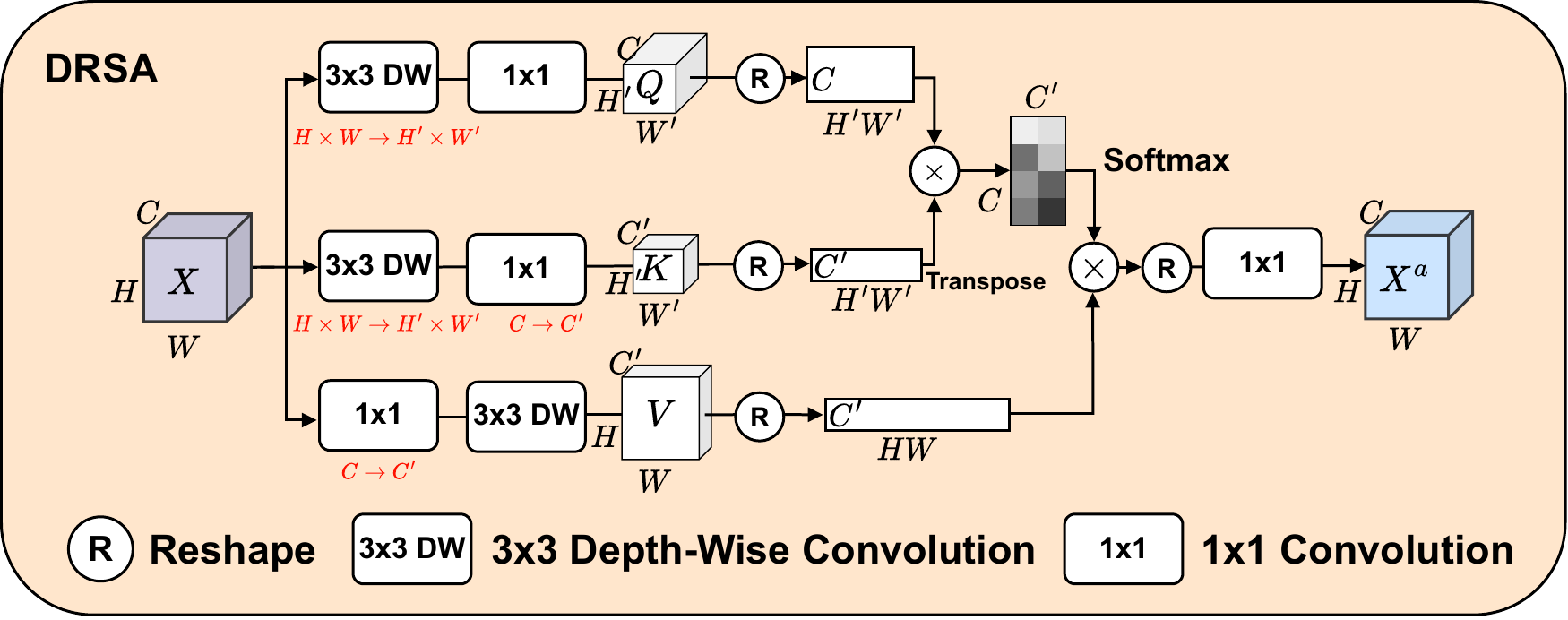} }
            {\scriptsize (c)}
        \end{minipage}
        \begin{minipage}[t]{0.14 \textwidth}
            \centering
            \raisebox{-2cm}{\includegraphics[width=1 \textwidth]{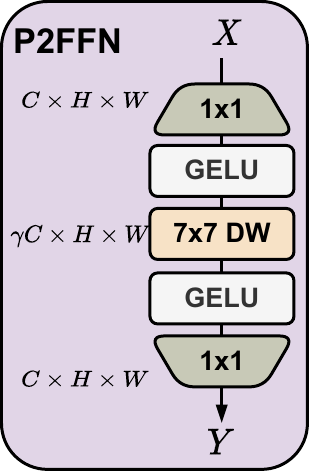} }
            {\scriptsize (d)}
        \end{minipage}
    \end{subfigure}
    
    \caption{Illustration of the proposed MARformer (a), the Transformer Block in our MARformer (b), the proposed Dimension-Reduced Self-Attention (DRSA) module (c), and the proposed Patch-wise Perceptive Feed Forward Network (P2FFN) (d).}
    \label{fig:marformer}
\end{figure*}

During the past decades, many metal artifact reduction (MAR) methods are developed, which can be roughly categorized into sinogram completion methods~\citep{kalender1987reduction,zhao2002wavelet,meyer2010normalized,zhang2011new}, optimization based methods~\citep{wang1996iterative,zhang2011metal,lemmens2008suppression}, and data driven methods~\citep{wang2018conditional,liao2019adn,ghani2019fast,zhang2018convolutional,lin2019dudonet,wang2021dual}. Sinogram completion methods mainly project CBCT images into the sinogram domain and then complete the corresponding metal influenced area via interpolation~\citep{kalender1987reduction,zhao2002wavelet,meyer2010normalized,zhang2011new}. Optimization based methods~\citep{wang1996iterative,zhang2011metal,lemmens2008suppression} model the MAR process as proper objective functions and reconstruct the clean CBCT images by optimization algorithms like ADMM~\citep{boyd2011distributed}. Data driven methods usually perform MAR by learning deep neural networks on paired clean and MA degraded CBCT images~\citep{wang2018conditional,liao2019adn,ghani2019fast,zhang2018convolutional,lin2019dudonet,wang2021dual}.

Despite achieving promising MAR performance, most existing deep MAR networks built upon convolutional neural networks (CNNs) are usually difficult to capture global correlation of metal artifacts in a CBCT image. A natural alternative is to develop Transformers for better MAR performance. By making better usage of global dependency~\citep{vaswani2017attention}, Transformers have outperformed CNNs on many tasks ranging from pattern recognition~\citep{dosovitskiy2020image} to image restoration~\citep{Zamir2021Restormer,wang2022uformer}. However, Transformer based models are usually parameter-demanding and computationally expensive~\citep{dosovitskiy2020image,xu2021vitae,Zamir2021Restormer}. To alleviate these problems, the token-wise learning scheme is developed in ViT~\citep{dosovitskiy2020image}. But this also reduces the correlation field of self-attention from global image to local patches, limiting the MAR performance. To this end, it is essential to develop an efficient and light-weight Transformer model that is feasible to exploit global correlation information for effective CBCT MAR.

In this paper, we develop an efficient Transformer to explore global information for effective metal artifact reduction of dental CBCT images. Since dental CBCT images are spatially similar in local areas, it is possible to alleviate the computational burden by performing multi-head self-attention (MHSA), which consumes the most computational costs in Transformers, on dimension-reduced feature tensors. To this end, we propose an efficient Dimension-Reduced Self-Attention (DRSA) module to reduce the computational complexity of standard MHSA, while maintaining its capability on exploiting global correlation by computing the similarity matrix along the channel dimension instead of spatial dimension in standard MHSA.  What’s more, after each DRSA module, we introduce a new Patch-wise Perceptive Feed Forward Network (P2FFN) module with a large convolution kernel to well recover the local details in dental CBCT images. Experiments on a large-scale dataset containing simulated and real-world dental CBCT images show that the proposed MAR Transformer (MARformer) outperforms previous MAR CNNs and image restoration Transformers.

\section{Related Work}
\label{sec:relatedwork}

% %-------------------------------------------------------------------------

\subsection{Metal Artifact Reduction}

Existing metal artifact reduction methods could be categorized into traditional sinogram completion, iterative reconstruction, and deep learning based medical image restoration.
Early sinogram completion methods rely on interpolating the missing information caused by metal objects in the sinogram field, using techniques like linear interpolation~\citep{kalender1987reduction}, wavelet interpolation~\citep{zhao2002wavelet}, and normalized interpolation~\citep{meyer2010normalized, zhang2011new}.
Iterative reconstruction techniques~\citep{wang1996iterative, wang1999iterative, zhang2011metal, lemmens2008suppression, mehranian2013x, chang2018prior} formulate MAR as an inverse problem, alternately enforcing data consistency and applying regularizers to reduce artifacts in the image domain.

With the rise of deep learning, several convolutional neural network (CNN) models have been proposed for MAR. Some operate purely in the image domain~\citep{xu2018deep, wang2018conditional}, while others complete the sinogram before reconstruction~\citep{ghani2019fast} or use dual domain approaches~\citep{zhang2018convolutional, lin2019dudonet, wang2021dual, wang2021indudonet}.
Most existing deep MAR methods use CNN architectures. Recently, transformers have shown promising effect for restoration tasks~\citep{swinIR9607618, wang2022uformer, Zamir2021Restormer}, but the high computation occupation of the model blocks their application. To overcome this problem, we employs a light-weight transformer model that has the similar computation cost with CNN models and shows remarkable performance on MAR task.

\subsection{Efficient Vision Transformers}

% % 介绍如何降低计算复杂度
% % transformer and  efficient transformer
Transformer has achieved great progress in natural language processing~\citep{vaswani2017attention} and computer vision~\citep{dosovitskiy2020image} for its ability to capture long range dependencies. Several works have enabled transformers to process images using multi-scale feature levels~\citep{9710580swintrans, wang2021pyramid}. However, a major drawback is their large model size and high computational cost. Some methods have aimed to reduce the parameters and computation of transformer~\citep{graham2021levit, hassani2021escaping, mehta2021mobilevit}.

Recently, transformers have been applied to image restoration tasks~\citep{swinIR9607618, Zamir2021Restormer, wang2022uformer}. \citep{swinIR9607618} utilizes the Swin Transformer~\citep{9710580swintrans} for restoration. Restormer~\citep{Zamir2021Restormer} and Uformer~\citep{wang2022uformer} employ U-Net like structures and demonstrate efficient computation for image restoration task. Vision transformers mainly compute self-attention across spatial dimensions, incurring high cost due to the quadratic computation complexity of the similarity matrix in the self-attention part. Similar to~\citep{Zamir2021Restormer} which is able to handle high resolution image, we compute self-attention across channels, since the number of channels is less than spatial size for restoration task. Furthermore, we introduce dimension reduction method in the transformer block to cut computation costs in both channel and spatial dimensions.

%% benchmark comparison
\begin{table*}[ht]
	\centering 
	\caption{\textbf{Quantitative results of PSNR (dB), SSIM~\citep{wang2004image} and Dice score over the synthesized MA dataset}. The FLOPs and speed are test on a sinle $400\times400$ image. ``-'' means that the result is not available. ``*'' means that  the method is test on an Intel Xeon Gold 6348 CPU. The best results are highlighted in \textbf{bold}.
	}
% 	\resizebox{1.0\textwidth}{!}{
		% \renewcommand\arraystretch{1.2}
		\begin{tabular}{r||ccc|ccc}
			\Xhline{1pt}

			\rowcolor[rgb]{.9, .9, .9}  
			Method
			& PSNR  & SSIM & Dice
			& Params(M) & FLOPs(G) & Speed(ms)
			\\
			\hline 
			Input MA Images\ 
			& 25.72 & 0.7207 & 0.6443
			  & - &-&-
			\\
			\hline
			LI%~\citep{kalender1987reduction}
			& 28.66 & 0.8822 & 0.6868
			  & - & - & 75*
			%\\NMAR1~\citep{meyer2010normalized}& 26.68 & 0.8729 & 0.6390 & -- & -- & 187
			\\
			NMAR%~\citep{meyer2010normalized}
			& 28.93 & 0.9023 & 0.6963
			  & - & - & 269*
			\\
			CNNMAR%~\citep{zhang2018convolutional}
			& 30.05 & 0.9391 & 0.7511
			  & - & - & 373*
			\\
			UNet-MAR%~\citep{ronneberger2015u}
			& 40.20 & 0.9707 & 0.7983
			& 12.25 & 44.28 & \textbf{10}
			\\  
			ResNet-MAR%~\citep{ledig2017photo}
			& 37.34 & 0.9626 & 0.7744
			& 0.59 & 95.34 & 18
			\\
			RDN-CT%~\citep{lin2019dudonet}
			& 39.68 & 0.9711 & 0.7929
			& 13.76 & 2200.88 & 256
			\\
			\hline	
			Restormer%~\citep{Zamir2021Restormer}
			& 42.73 & 0.9777 & 0.8014
			& 26.12 & 377.72 & 208
			\\	Uformer-B%~\citep{wang2022uformer}
			& 42.98 & \textbf{0.9790} & \textbf{0.8041}
			& 50.42 & 205.82  & 116
			\\	Uformer-S%~\citep{wang2022uformer}
			& 42.35 & 0.9770 & 0.8039
			& 20.66 & 95.84  & 90
			\\	Uformer-T%~\citep{wang2022uformer}
			& 41.44 & 0.9744 & 0.8024
			& 5.24 & 25.39 & 51
			\\
			\hline
			\textbf{MARformer-L} 
			& \textbf{43.11} & 0.9789 & 0.8031
			& 11.76  & 60.25 & 48
			\\
			\textbf{MARformer-B} 
			& 42.89 & 0.9782 & 0.8021
			& 6.88 & 46.20 & 36
			\\
			\textbf{MARformer-T} 
			& 41.40 & 0.9736 & 0.8011
			& \textbf{0.40} & \textbf{12.82}  & 31
			\\
			\hline
		\end{tabular}
% 	}
	\label{tab:benchmark_pairedma}
\end{table*}

\section{Proposed MARformer for Dental CBCT Image MAR}

\subsection{Network Overview}

As shown in \cref{fig:marformer} , the proposed MARformer utilizes a U-net architecture with three-level hierarchial encoder-decoders and the $i$-th ($i=1,2,3$) level of encoder-decoder contains $N_{i}$ transformer blocks, together with a bottleneck including $N_4$ transformer blocks. Each transformer block is consisted of the proposed Dimension-Reduced Self-Attention (DRSA) module and the Patch-wise Perceptive Feed Forward Network (P2FFN), as shown in \cref{fig:marformer} (b).

Given an input dental CBCT image degraded by metal artifacts $\mathbf{I}\in\mathbb{R}^{H\times W}$, we first extract initial image feature $\mathbf{X}_0\in\mathbb{R}^{C\times H\times W}$ by a 3$\times$3 convolution. Then the image feature $\mathbf{X}_0$ is fed into the $1$-st level encoder. The output feature is spatial-wise downsampled and channel-wise expanded to a size of $2C\times\frac{H}{2}\times\frac{W}{2}$. The resulting feature map is further fed into the $2$-nd level encoder. After three levels of hierarchial encoders, the output feature map of size $8C\times\frac{H}{8}\times\frac{W}{8}$ is input to the bottleneck.
In the decoder stage, the feature map is spatial-wise upsampled and channel-wise shrinked by half in each level of decoder, and concatenated with the feature output by the corresponding-level encoder. To keep the channel dimension consistent between different levels, we use a 1$\times$1 convolution to reduce the dimension of the concatenated feature from 2$C$ to $C$ before it is fed into the next-level decoder. The downsample and upsample operations between different levels of encoder or decoders are implemented by pixel unshuffle/shuffle operations~\citep{shi2016real}. The feature map after the last decoder is fused by a 3$\times$3 convolution to get the image residual $\mathbf{R}\in\mathbb{R}^{H\times W}$, which is added to the input image $\mathbf{I}$ to produce the final MAR image $\hat{\mathbf{I}}=\mathbf{I}+\mathbf{R}$.

\subsection{Components of the Transformer Block}
In our MARformer, each transformer block is consisted of a Dimension-Reduced Self-Attention (DRSA) module and a Patch-wise Perceptive Feed Forward Network (P2FFN), as shown in \cref{fig:marformer} (b).

%------------------------------------
\noindent
\textbf{Dimension Reduced Self-Attention (DRSA)}.
The goal of the proposed DRSA module is to perform efficient self-attention on high-resolution feature maps. Given an input feature $\mathbf{X}\in\mathbb{R}^{C\times H\times W}$, we compute global similarity matrix~\citep{vaswani2017attention,dosovitskiy2020image} along the channel dimension, instead of spatial dimension~\citep{vaswani2017attention,dosovitskiy2020image}. Assume that there is only one head of self-attention, our DRSA module first projects the input feature $\mathbf{X}$ into spatially-reduced query component $\mathbf{Q}\in\mathbb{R}^{C\times H'W'}$ ($H'<H$, $W'<W$), key component $\mathbf{K}\in\mathbb{R}^{C'\times HW}$, and value component $\mathbf{V}\in\mathbb{R}^{C'\times HW}$ ($C'<C$). The projection is implemented by a 1$\times$1 convolution and a 3$\times$3 depth-wise convolution, as plotted in~\cref{fig:marformer} (c). For example, the spatial downsampling is realized by 3$\times$3 convolutions with a stride larger than 1 (e.g., 2) while the channel downsampling is realized by 1$\times$1 convolutions. Then the DRSA is performed as $\text{Attention} = \text{Softmax}(\frac{1}{\alpha}\mathbf{Q}\mathbf{K}^T)\mathbf{V}$, where ``\text{Softmax}’’ is the softmax normalization operation and $\alpha$ is a learnable scalar parameter.

In our DRSA, since the image size $HW$ is usually much larger than the channel dimension $C$, the computational complexity of calculating the similarity matrix along the channel dimension is $\mathcal{O}(CC' H'W' + CC' HW)$, which is smaller than the computational complexity by calculating spatial-wise similarity matrix, i.e., $\mathcal{O}(C(HW)^2)$, when projecting the input feature $\mathbf{X}$ onto feature maps of $\mathbf{Q}, \mathbf{K}, \mathbf{V}\in\mathbb{R}^{HW\times C}$. Besides, calculating channel-wise similarity matrix is also able to enable our MARformer to exploit global correlation information for effective dental CBCT image MAR. We also split the channel dimension into multiple groups to perform multi-head self-attention.

\noindent
\textbf{Patch-wise Perceptive Feed Forward Network (P2FFN)}.
As is shown in~\cref{fig:marformer} (d), our P2FFN is consisted of a 1$\times$1 convolution to expand the channel dimension from $C$ to $\gamma C$, a GELU activation function~\citep{hendrycks2016gelu}, a depth-wise $p\times p$ convolution with a GELU to perceive local image information that is important to image restoration tasks, and another 1$\times$1 convolution to shrink the channel dimension from $\gamma C$ to $C$. Finally, we add a skip connection with the input feature to the end of P2FFN to improve the learning capability of our MARformer.

\subsection{Implementation details}

The $k$-th level ($k=1,2,3$) of the encoder-decoder in our MARformer contains $N_k$ transformer blocks, and the bottleneck contains $N_4$ transformer blocks. The numbers of heads when calculating the similarity matrices in transformer blocks at different levels of encoder-decoders are denoted as $H_1, H_2, H_3, H_4$. The spatial and channel downsample ratio $r$ are both set as 2. We set the channel dimension $C=48$ and the expansion factor $\gamma=2$ in P2FFN.

We implement three models of our MARformer with different parameter amounts. For the largest one, denoted as MARformer-L, we set $N_1=1$, $N_2=2$, $N_3=4$, $N_4=8$ and $H_1=1$, $H_2=2$, $H_3=4$, $H_4=8$. For the baseline one, denoted as MARformer-B, we set $N_1=1$, $N_2=2$, $N_3=3$, $N_4=4$ and $H_1=1$, $H_2=2$, $H_3=4$, $H_4=8$. For the tiny one, denoted as MARformer-T, we set $N_1=1$, $N_2=2$, $N_3=3$, $N_4=4$, $H_1=H_2=H_3=H_4=1$, and fix the channel dimension as $48$ in all levels of encoder-decoders and the bottleneck.

%%synthetic image
\begin{figure*}[p]
	\centering
	% first image
         \setstretch{0.8}
	\begin{subfigure}
		\centering
		\begin{minipage}[t]{0.18\textwidth}
			\centering % 1a
			\raisebox{-0.1cm}{\includegraphics[width=1\textwidth]{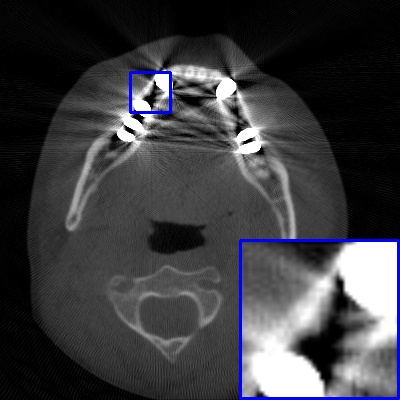}} 
			{\scriptsize MA \\ 22.38 / 0.6196}
		\end{minipage}
		\begin{minipage}[t]{0.18\textwidth}
			\centering % (2-a)
			\raisebox{-0.1cm}{\includegraphics[width=1\textwidth]{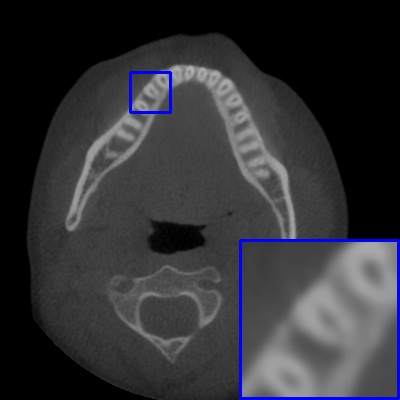} }
			{\scriptsize  UNet-MAR \\ 41.60 / 0.9752 }
		\end{minipage}
		\begin{minipage}[t]{0.18\textwidth}
			\centering % 1a
			\raisebox{-0.1cm}{\includegraphics[width=1\textwidth]{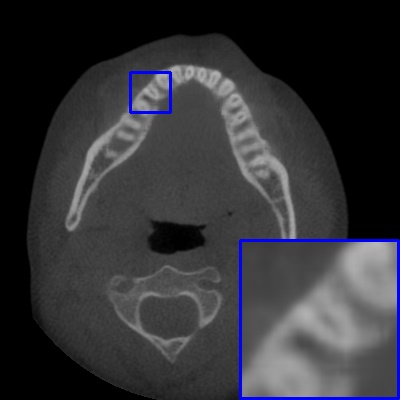}} 
			{\scriptsize  ResNet-MAR\\37.48 / 0.9663}
		\end{minipage}
		\begin{minipage}[t]{0.18\textwidth}
			\centering % 1b
			\raisebox{-0.1cm}{\includegraphics[width=1\textwidth]{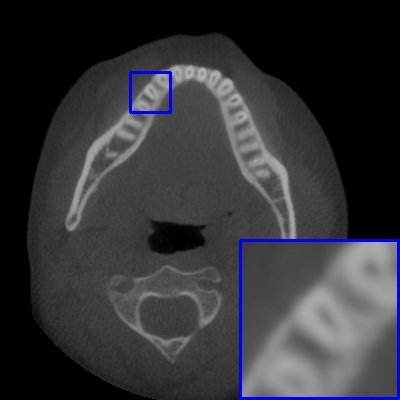}}
			{\scriptsize  RDN-CT \\ 40.55 / 0.9745}
		\end{minipage}
		\begin{minipage}[t]{0.18\textwidth}
			\centering %1c
			\raisebox{-0.1cm}{\includegraphics[width=1\textwidth]{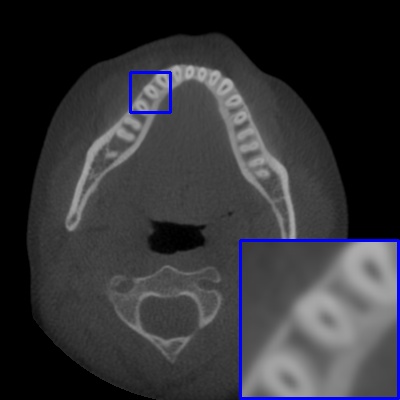}}
			{\scriptsize  Restormer\\43.19 / 0.9789 }
		\end{minipage}
	\end{subfigure}
	\begin{subfigure}
		\centering
		\begin{minipage}[t]{0.18\textwidth}
			\centering %1c
			\raisebox{-0.1cm}{\includegraphics[width=1\textwidth]{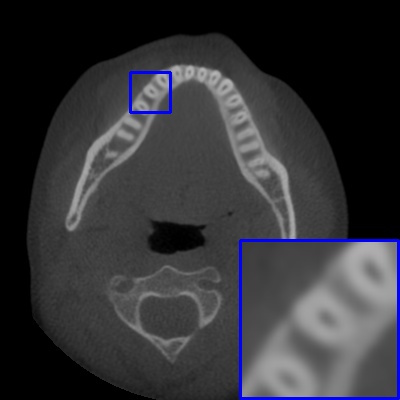}}
			{\scriptsize  Uformer-T \\ 41.92 / 0.9761}
		\end{minipage}
		\begin{minipage}[t]{0.18\textwidth}
			\centering %1c
			\raisebox{-0.1cm}{\includegraphics[width=1\textwidth]{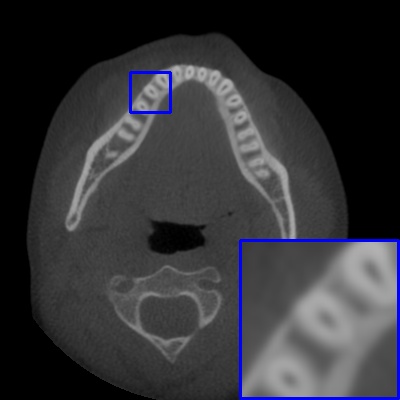}}
			{\scriptsize  Uformer-B \\ 43.43 / 0.9801}
		\end{minipage}
		\begin{minipage}[t]{0.18\textwidth}
			\centering %1c
			\raisebox{-0.1cm}{\includegraphics[width=1\textwidth]{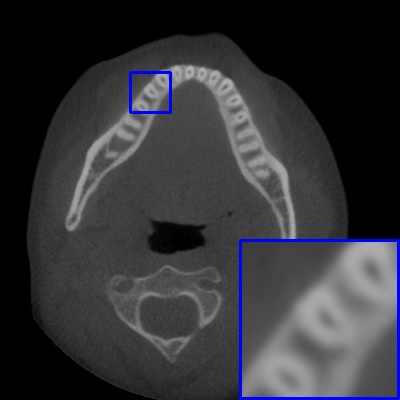}}
			{\scriptsize  MARformer-T \\ 41.98 / 0.9757}
		\end{minipage}
		\begin{minipage}[t]{0.18\textwidth}
			\centering %1c
			\raisebox{-0.1cm}{\includegraphics[width=1\textwidth]{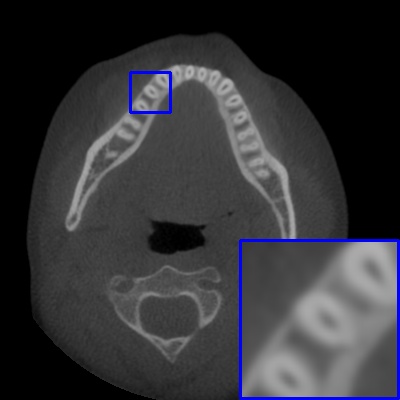}}
			{\scriptsize  MARformer-L \\ 43.71 / 0.9803}
		\end{minipage}
		\begin{minipage}[t]{0.18\textwidth}
			\centering % (2-a)
			\raisebox{-0.1cm}{\includegraphics[width=1\textwidth]{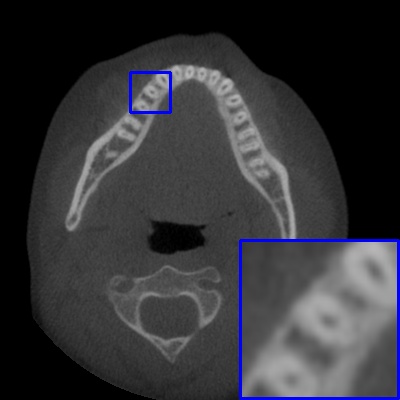} }
			{\scriptsize  GT\\ $\infty$ / 1.0000}
		\end{minipage}
	\end{subfigure}
        \vspace{0.5mm}
        \hrule
        \vspace{0.5mm}

	\begin{subfigure}
		\centering
		\begin{minipage}[t]{0.18\textwidth}
			\centering % 1a
			\raisebox{-0.1cm}{\includegraphics[width=1\textwidth]{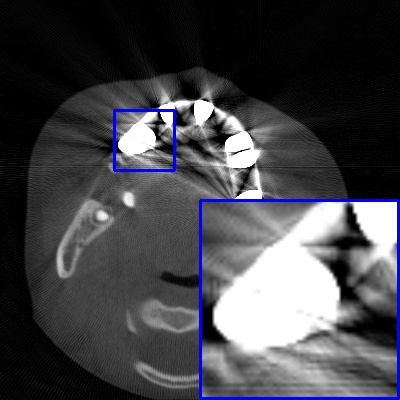}} 
			{\scriptsize  MA \\ 20.48 / 0.5851}
		\end{minipage}
		\begin{minipage}[t]{0.18\textwidth}
			\centering % (2-a)
			\raisebox{-0.1cm}{\includegraphics[width=1\textwidth]{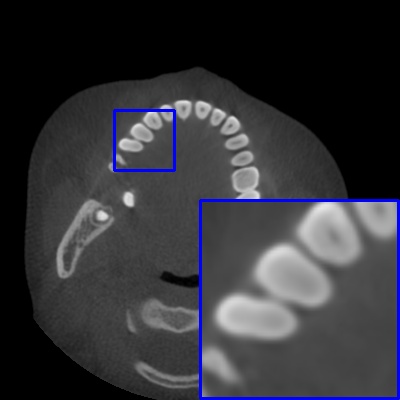} }
			{\scriptsize  UNet-MAR \\ 35.69 / 0.9614}
		\end{minipage}
		\begin{minipage}[t]{0.18\textwidth}
			\centering % 1a
			\raisebox{-0.1cm}{\includegraphics[width=1\textwidth]{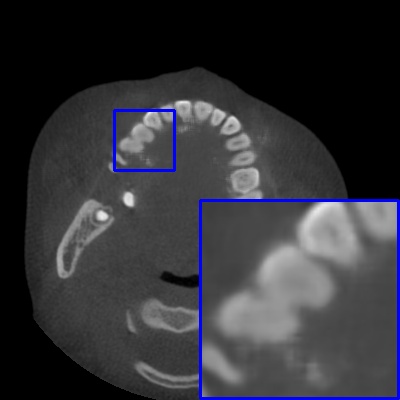}} 
			{\scriptsize  ResNet-MAR \\ 33.13 / 0.9505 }
		\end{minipage}
		\begin{minipage}[t]{0.18\textwidth}
			\centering % 1b
			\raisebox{-0.1cm}{\includegraphics[width=1\textwidth]{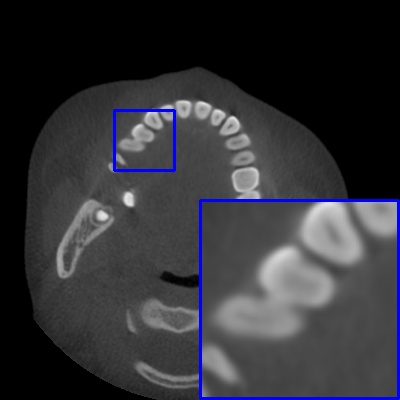}}
			{\scriptsize   RDN-CT \\ 35.50 / 0.9623 }
		\end{minipage}
		\begin{minipage}[t]{0.18\textwidth}
			\centering %1c
			\raisebox{-0.1cm}{\includegraphics[width=1\textwidth]{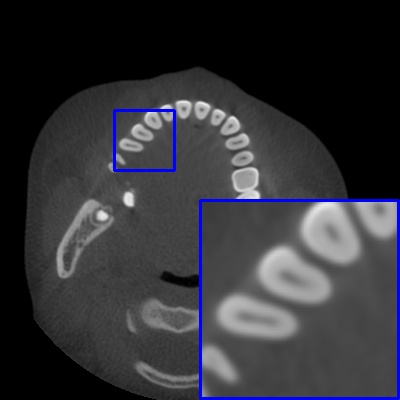}}
			{\scriptsize  Restormer \\ 39.43 / 0.9700 }
		\end{minipage}
	\end{subfigure}
	\begin{subfigure}
		\centering
		\begin{minipage}[t]{0.18\textwidth}
			\centering %1c
			\raisebox{-0.1cm}{\includegraphics[width=1\textwidth]{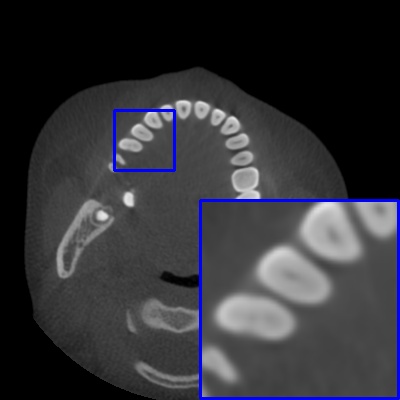}}
			{\scriptsize  Uformer-T \\ 37.52 / 0.9657 }
		\end{minipage}
		\begin{minipage}[t]{0.18\textwidth}
			\centering %1c
			\raisebox{-0.1cm}{\includegraphics[width=1\textwidth]{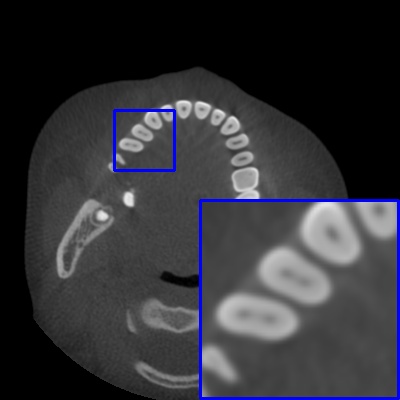}}
			{\scriptsize  Uformer-B \\ 38.30 / 0.9711 }
		\end{minipage}
		\begin{minipage}[t]{0.18\textwidth}
			\centering %1c
			\raisebox{-0.1cm}{\includegraphics[width=1\textwidth]{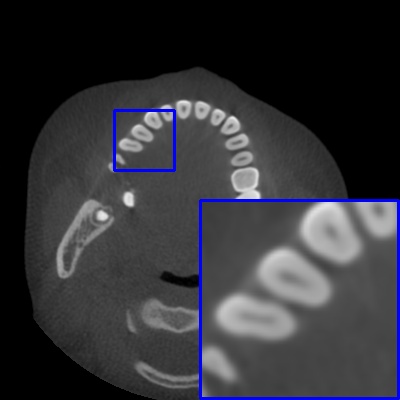}}
			{\scriptsize  MARformer-T \\ 37.72 / 0.9655 }
		\end{minipage}
		\begin{minipage}[t]{0.18\textwidth}
			\centering %1c
			\raisebox{-0.1cm}{\includegraphics[width=1\textwidth]{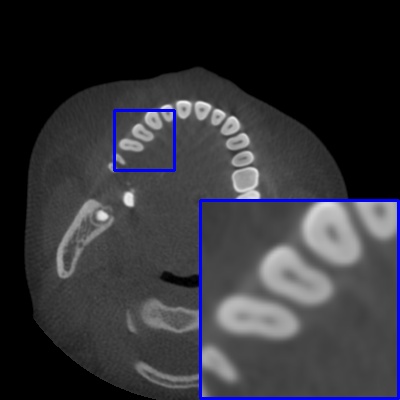}}
			{\scriptsize MARformer-L \\ \textbf{39.97} / \textbf{0.9726} }
		\end{minipage}
		\begin{minipage}[t]{0.18\textwidth}
			\centering % (2-a)
			\raisebox{-0.1cm}{\includegraphics[width=1\textwidth]{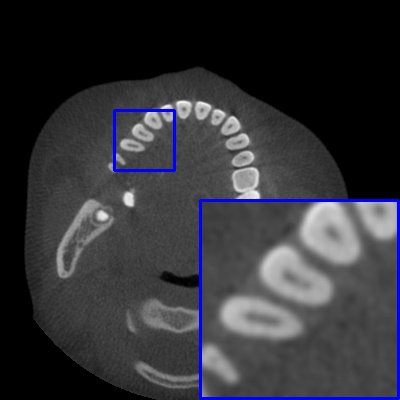} }
			{\scriptsize  GT \\ $\infty$ / 1.0000}
		\end{minipage}
	\end{subfigure}
        \vspace{0.5mm}
        \hrule
        \vspace{0.5mm}
        % \vspace{0.6mm}

 	\begin{subfigure}
		\centering
		\begin{minipage}[t]{0.18\textwidth}
			\centering % 1a
			\raisebox{-0.1cm}{\includegraphics[width=1\textwidth]{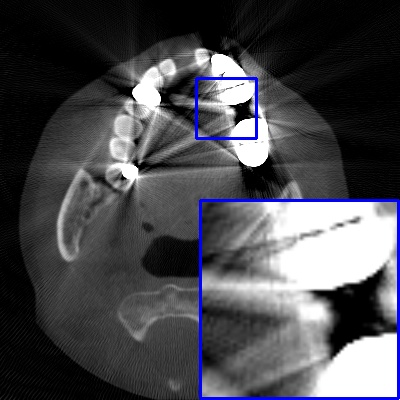}} 
			{\scriptsize  MA \\ 19.37 / 0.5517 }
		\end{minipage}
		\begin{minipage}[t]{0.18\textwidth}
			\centering % (2-a)
			\raisebox{-0.1cm}{\includegraphics[width=1\textwidth]{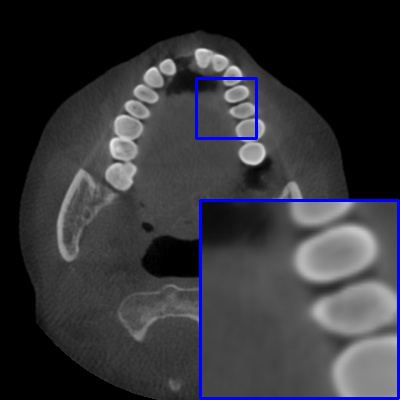} }
			{\scriptsize  UNet-MAR \\ 34.31 / 0.9535 }
		\end{minipage}
		\begin{minipage}[t]{0.18\textwidth}
			\centering % 1a
			\raisebox{-0.1cm}{\includegraphics[width=1\textwidth]{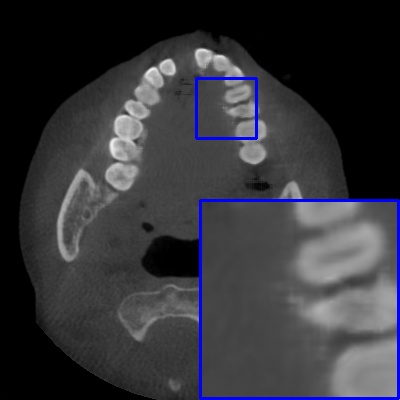}} 
			{\scriptsize  ResNet-MAR \\ 29.94 / 0.9302 }
		\end{minipage}
		\begin{minipage}[t]{0.18\textwidth}
			\centering % 1b
			\raisebox{-0.1cm}{\includegraphics[width=1\textwidth]{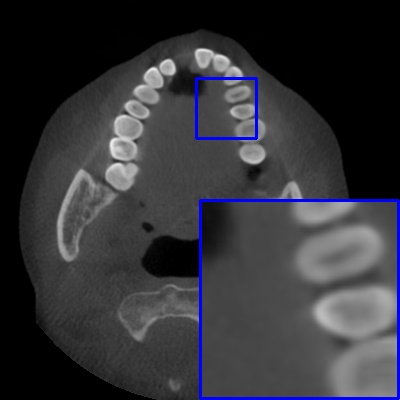}}
			{\scriptsize   RDN-CT \\ 31.87 / 0.9488 }
		\end{minipage}
		\begin{minipage}[t]{0.18\textwidth}
			\centering %1c
			\raisebox{-0.1cm}{\includegraphics[width=1\textwidth]{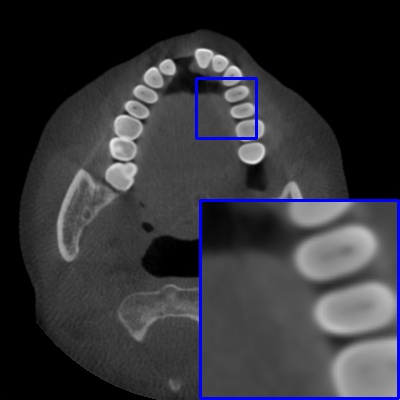}}
			{\scriptsize  Restormer \\ 35.88 / 0.9652 }
		\end{minipage}
	\end{subfigure}
	\begin{subfigure}
		\centering
		\begin{minipage}[t]{0.18\textwidth}
			\centering %1c
			\raisebox{-0.1cm}{\includegraphics[width=1\textwidth]{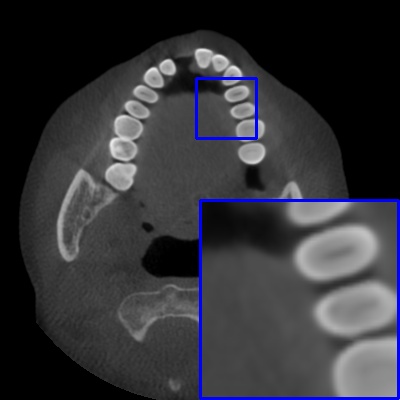}}
			{\scriptsize  Uformer-T \\ 35.17 / 0.9604 }
		\end{minipage}
		\begin{minipage}[t]{0.18\textwidth}
			\centering %1c
			\raisebox{-0.1cm}{\includegraphics[width=1\textwidth]{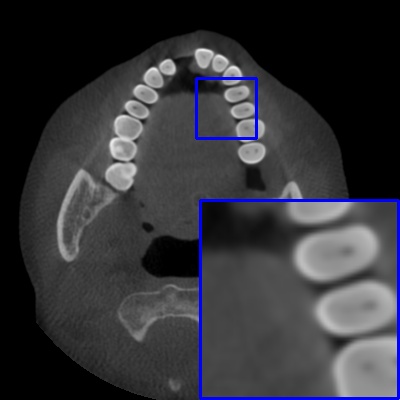}}
			{\scriptsize  Uformer-B \\ 35.85 / 0.9673 }
		\end{minipage}
		\begin{minipage}[t]{0.18\textwidth}
			\centering %1c
			\raisebox{-0.1cm}{\includegraphics[width=1\textwidth]{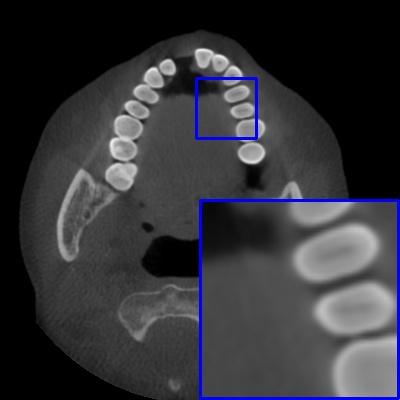}}
			{\scriptsize  MARformer-T \\ 35.22 / 0.9588 }
		\end{minipage}
		\begin{minipage}[t]{0.18\textwidth}
			\centering %1c
			\raisebox{-0.1cm}{\includegraphics[width=1\textwidth]{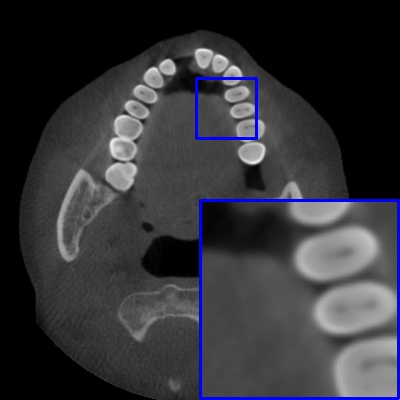}}
			{\scriptsize  MARformer-L \\ \textbf{37.37} / \textbf{0.9678}}
		\end{minipage}
		\begin{minipage}[t]{0.18\textwidth}
			\centering % (2-a)
			\raisebox{-0.1cm}{\includegraphics[width=1\textwidth]{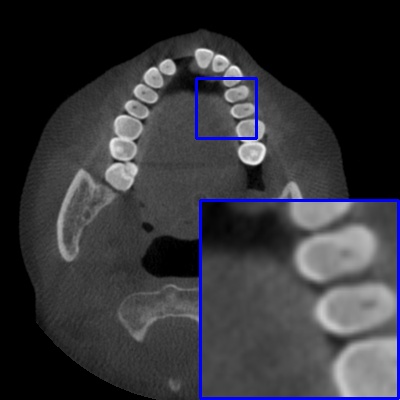} }
			{\scriptsize  GT \\ $\infty$ / 1.0000}
		\end{minipage}
	\end{subfigure}
 
	\caption{Visual comparison of MAR images by different methods on synthetic MA image. The PSNR (dB)/SSIM results are reported below each image for reference.}
	% Evaluation metric listed below each image is PSNR/SSIM.
	\label{visual-group1}
\end{figure*}

%%%% Clinical Images
\begin{figure*}[p]
	\centering
	
	%% 4-th image
	\begin{subfigure}
		\centering
		\begin{minipage}[t]{0.18\textwidth}
			\centering % 1a
			\raisebox{-0.1cm}{\includegraphics[width=1\textwidth]{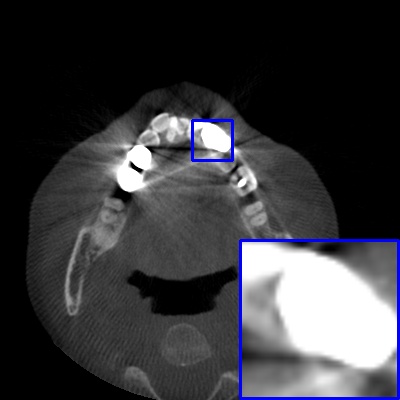}} 
			{\scriptsize  MA}
		\end{minipage}
		\begin{minipage}[t]{0.18\textwidth}
			\centering % (2-a)
			\raisebox{-0.1cm}{\includegraphics[width=1\textwidth]{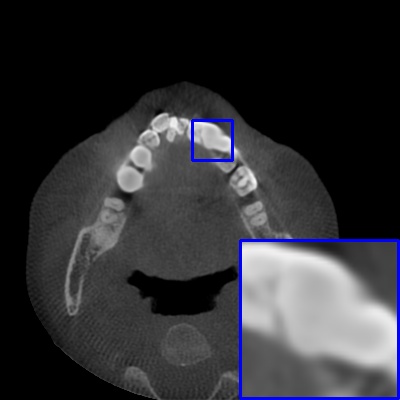} }
			{\scriptsize  UNet-MAR }
		\end{minipage}
		\begin{minipage}[t]{0.18\textwidth}
			\centering % 1a
			\raisebox{-0.1cm}{\includegraphics[width=1\textwidth]{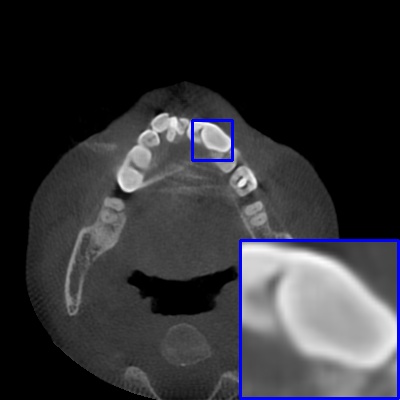}} 
			{\scriptsize  ResNet-MAR }
		\end{minipage}
		\begin{minipage}[t]{0.18\textwidth}
			\centering % 1b
			\raisebox{-0.1cm}{\includegraphics[width=1\textwidth]{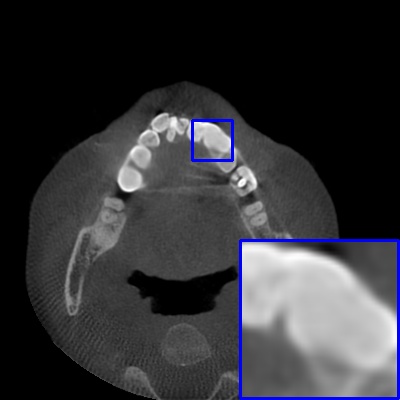}}
			{\scriptsize  RDN-CT}
		\end{minipage}
		\begin{minipage}[t]{0.18\textwidth}
			\centering %1c
			\raisebox{-0.1cm}{\includegraphics[width=1\textwidth]{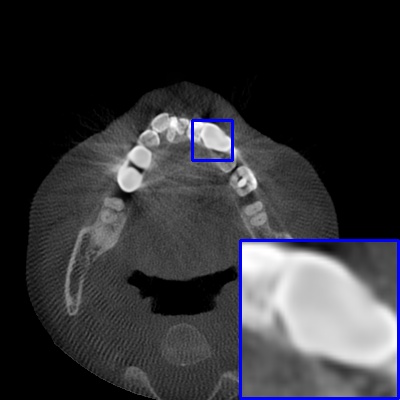}}
			{\scriptsize  Restormer }
		\end{minipage}
	\end{subfigure}
	\begin{subfigure}
		\centering
		\begin{minipage}[t]{0.18\textwidth}
			\centering %1c
			\raisebox{-0.1cm}{\includegraphics[width=1\textwidth]{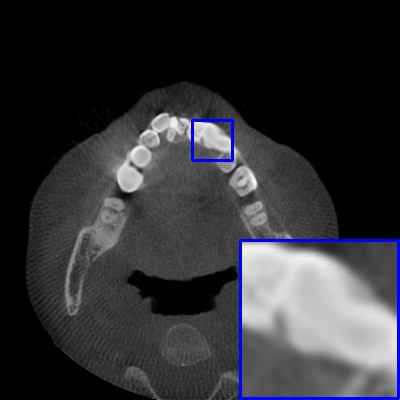}}
			{\scriptsize  Uformer-T }
		\end{minipage}
		\begin{minipage}[t]{0.18\textwidth}
			\centering %1c
			\raisebox{-0.1cm}{\includegraphics[width=1\textwidth]{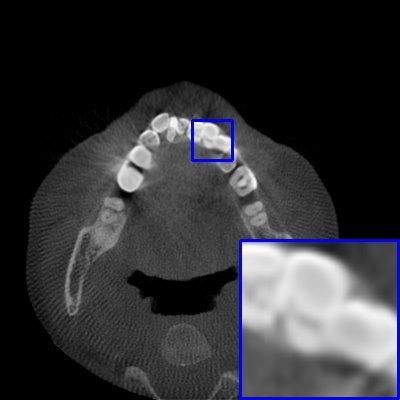}}
			{\scriptsize  Uformer-B }
		\end{minipage}
		\begin{minipage}[t]{0.18\textwidth}
			\centering %1c
			\raisebox{-0.1cm}{\includegraphics[width=1\textwidth]{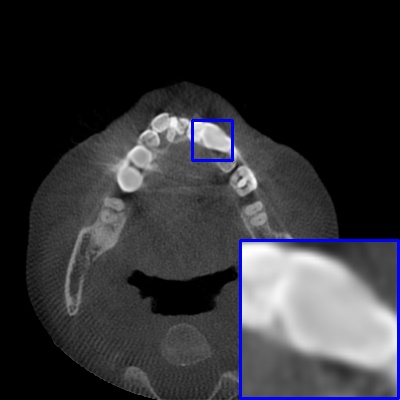}}
			{\scriptsize  MARformer-T}
		\end{minipage}
		\begin{minipage}[t]{0.18\textwidth}
			\centering %1c
			\raisebox{-0.1cm}{\includegraphics[width=1\textwidth]{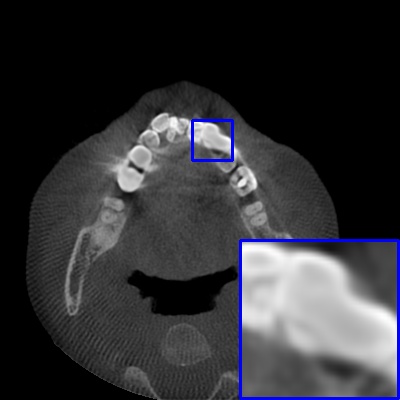}}
			{\scriptsize  MARformer-L}
		\end{minipage}
		\begin{minipage}[t]{0.18\textwidth}
			\centering %1c
			\raisebox{-0.1cm}{\includegraphics[width=1\textwidth]{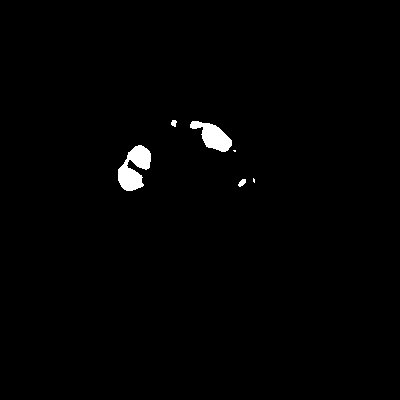}}
			{\scriptsize  Metal mask}
		\end{minipage}
	\end{subfigure}
        \vspace{0.5mm}
        \hrule
        \vspace{0.5mm}

	%% 5-th image
	\begin{subfigure}
		\centering
		\begin{minipage}[t]{0.18\textwidth}
			\centering % 1a
			\raisebox{-0.1cm}{\includegraphics[width=1\textwidth]{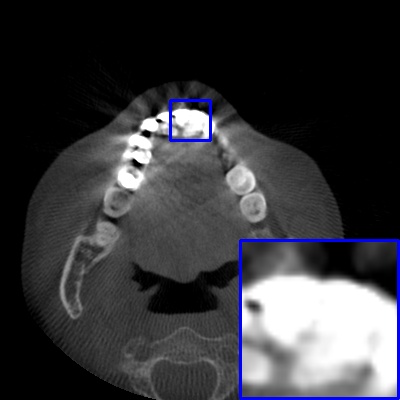}} 
			{\scriptsize  MA}
		\end{minipage}
		\begin{minipage}[t]{0.18\textwidth}
			\centering % (2-a)
			\raisebox{-0.1cm}{\includegraphics[width=1\textwidth]{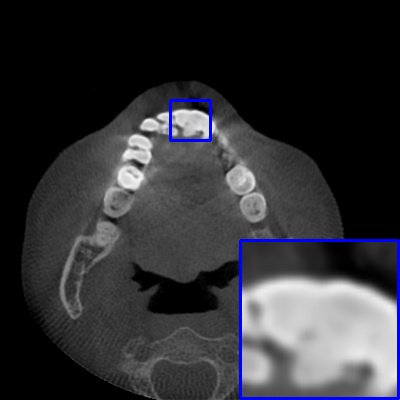} }
			{\scriptsize  UNet-MAR}
		\end{minipage}
		\begin{minipage}[t]{0.18\textwidth}
			\centering % 1a
			\raisebox{-0.1cm}{\includegraphics[width=1\textwidth]{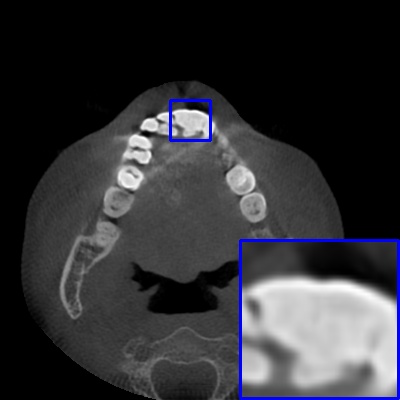}} 
			{\scriptsize  ResNet-MAR}
		\end{minipage}
		\begin{minipage}[t]{0.18\textwidth}
			\centering % 1b
			\raisebox{-0.1cm}{\includegraphics[width=1\textwidth]{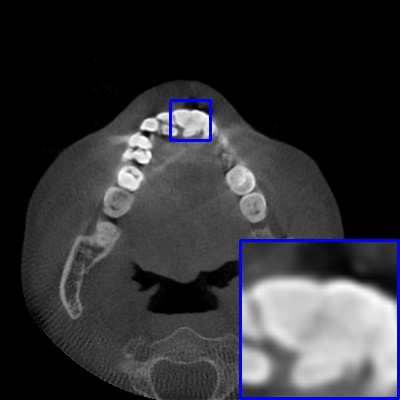}}
			{\scriptsize  RDN-CT}
		\end{minipage}
		\begin{minipage}[t]{0.18\textwidth}
			\centering %1c
			\raisebox{-0.1cm}{\includegraphics[width=1\textwidth]{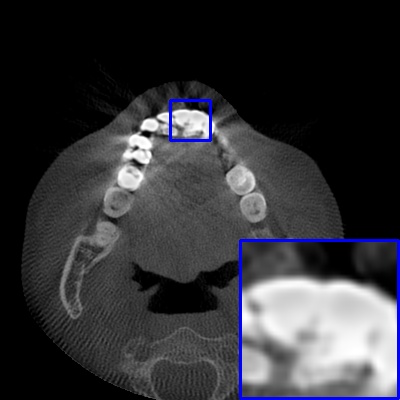}}
			{\scriptsize  Restormer }
		\end{minipage}
	\end{subfigure}
	\begin{subfigure}
		\centering
		\begin{minipage}[t]{0.18\textwidth}
			\centering %1c
			\raisebox{-0.1cm}{\includegraphics[width=1\textwidth]{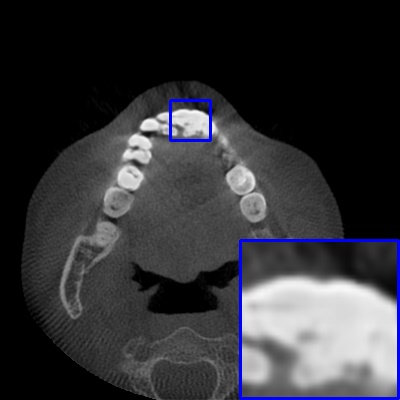}}
			{\scriptsize   Uformer-T }
		\end{minipage}
		\begin{minipage}[t]{0.18\textwidth}
			\centering %1c
			\raisebox{-0.1cm}{\includegraphics[width=1\textwidth]{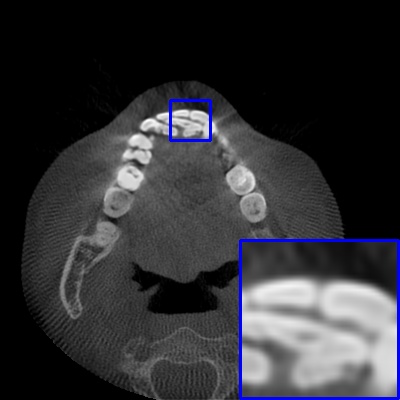}}
			{\scriptsize   Uformer-B }
		\end{minipage}
		\begin{minipage}[t]{0.18\textwidth}
			\centering %1c
			\raisebox{-0.1cm}{\includegraphics[width=1\textwidth]{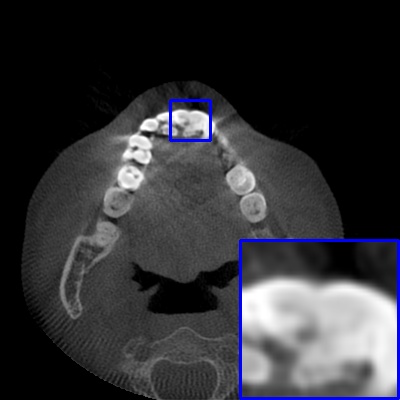}}
			{\scriptsize   MARformer-T }
		\end{minipage}
		\begin{minipage}[t]{0.18\textwidth}
			\centering %1c
			\raisebox{-0.1cm}{\includegraphics[width=1\textwidth]{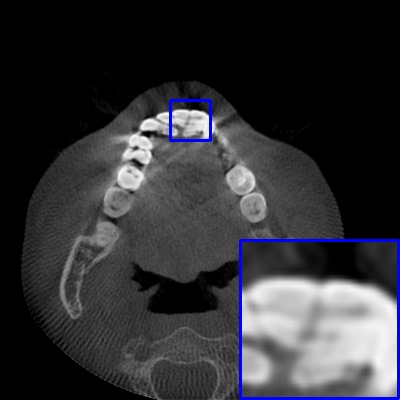}}
			{\scriptsize   MARformer-L}
		\end{minipage}
		\begin{minipage}[t]{0.18\textwidth}
			\centering %1c
			\raisebox{-0.1cm}{\includegraphics[width=1\textwidth]{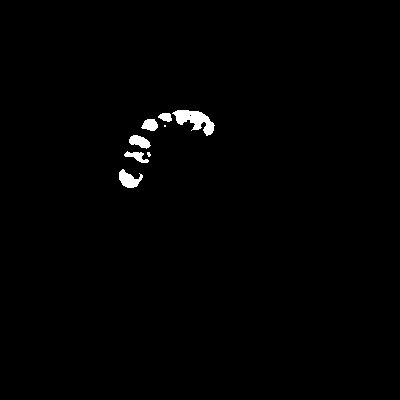}}
			{\scriptsize   Metal mask}
		\end{minipage}
	\end{subfigure}
        \vspace{0.5mm}
        \hrule
        \vspace{0.5mm}
        % \vspace{0.6mm}

	%% 6-th image
	\begin{subfigure}
		\centering
		\begin{minipage}[t]{0.18\textwidth}
			\centering % 1a
			\raisebox{-0.1cm}{\includegraphics[width=1\textwidth]{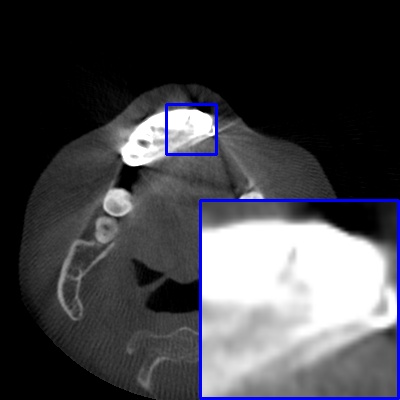}} 
			{\scriptsize  MA}
		\end{minipage}
		\begin{minipage}[t]{0.18\textwidth}
			\centering % (2-a)
			\raisebox{-0.1cm}{\includegraphics[width=1\textwidth]{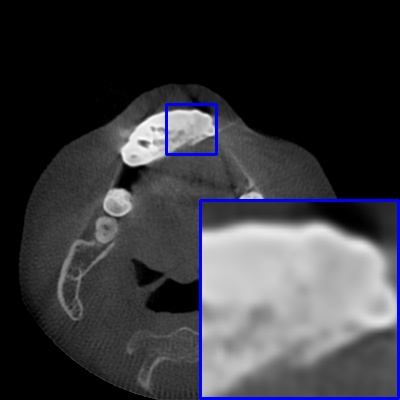} }
			{\scriptsize  UNet-MAR}
		\end{minipage}
		\begin{minipage}[t]{0.18\textwidth}
			\centering % 1a
			\raisebox{-0.1cm}{\includegraphics[width=1\textwidth]{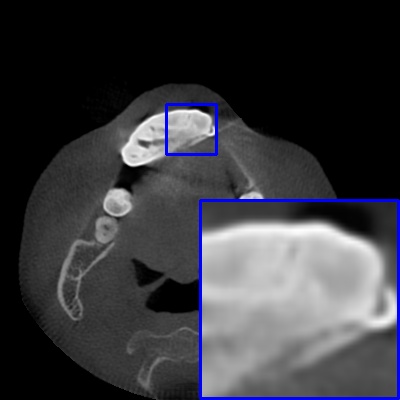}} 
			{\scriptsize  ResNet-MAR}
		\end{minipage}
		\begin{minipage}[t]{0.18\textwidth}
			\centering % 1b
			\raisebox{-0.1cm}{\includegraphics[width=1\textwidth]{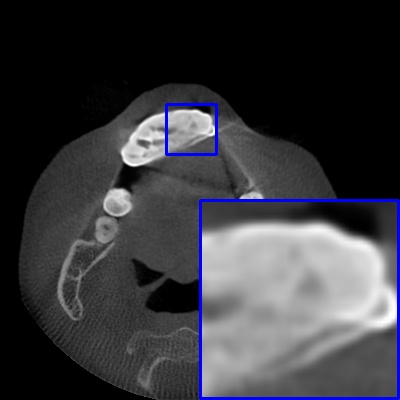}}
			{\scriptsize  RDN-CT}
		\end{minipage}
		\begin{minipage}[t]{0.18\textwidth}
			\centering %1c
			\raisebox{-0.1cm}{\includegraphics[width=1\textwidth]{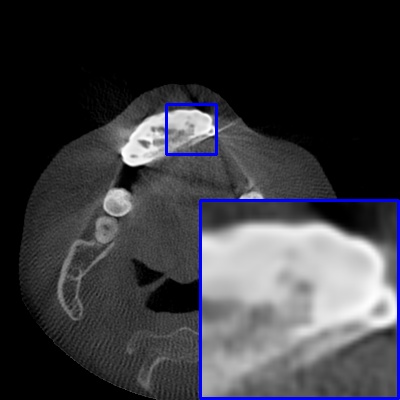}}
			{\scriptsize   Restormer}
		\end{minipage}
	\end{subfigure}
	
	\begin{subfigure}
		\centering
		\begin{minipage}[t]{0.18\textwidth}
			\centering %1c
			\raisebox{-0.1cm}{\includegraphics[width=1\textwidth]{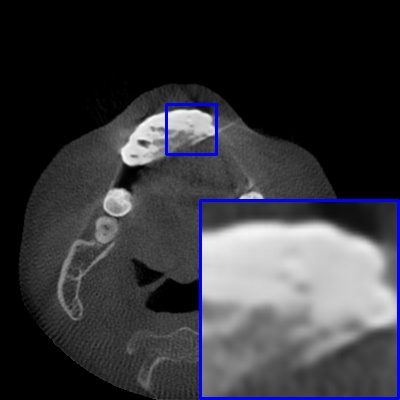}}
			{\scriptsize   Uformer-T }
		\end{minipage}
		\begin{minipage}[t]{0.18\textwidth}
			\centering %1c
			\raisebox{-0.1cm}{\includegraphics[width=1\textwidth]{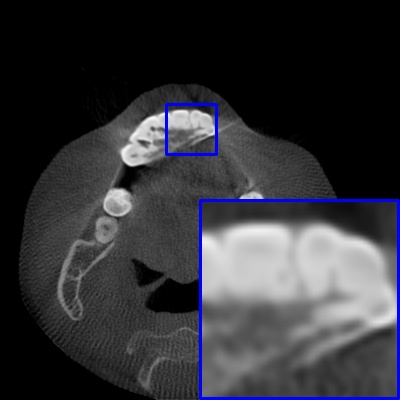}}
			{\scriptsize   Uformer-B }
		\end{minipage}
		\begin{minipage}[t]{0.18\textwidth}
			\centering %1c
			\raisebox{-0.1cm}{\includegraphics[width=1\textwidth]{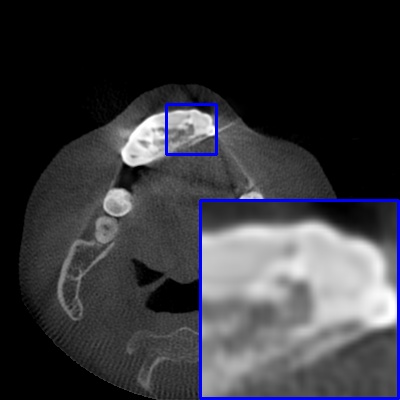}}
			{\scriptsize   MARformer-T }
		\end{minipage}
		\begin{minipage}[t]{0.18\textwidth}
			\centering %1c
			\raisebox{-0.1cm}{\includegraphics[width=1\textwidth]{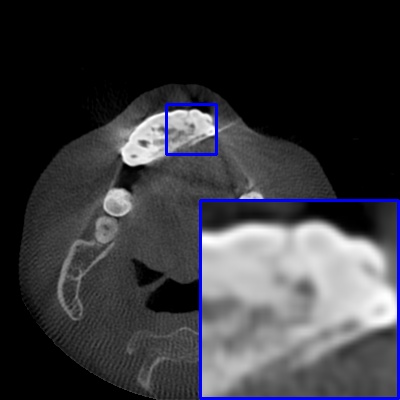}}
			{\scriptsize   MARformer-L }
		\end{minipage}
		\begin{minipage}[t]{0.18\textwidth}
			\centering % (2-a)
			\raisebox{-0.1cm}{\includegraphics[width=1\textwidth]{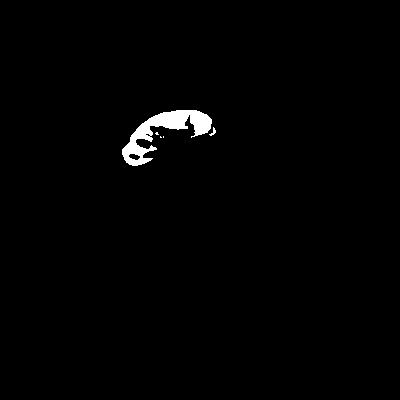} }
			{\scriptsize  Metal mask}
		\end{minipage}
	\end{subfigure}
 
	\caption{Comparison of MAR images by different methods on real-world MA image. The last image is the metal mask by selecting the pixel area over 2800HU in the MA image.
 % The last image is the segmentation mask by Poolformer on the result of MARformer-L.
 }
	\label{visual-clinical}
\end{figure*}

\section{Experimental Results}
\subsection{Dataset Preparation}
\label{sec:dataprepare}

To evaluate different methods on metal artifacts removal, we collected 263 CBCT scans, each of which contains $224$ to $448$ slice images at a resolution of $0.4\text{mm}\times 0.4\text{mm}$, $0.3\text{mm}\times 0.3\text{mm}$, or $0.25\text{mm}\times 0.25\text{mm}$ from over ten stomatological hospitals.
The values of CT slices are clipped between -1,000HU and 2,800HU, and then converted to attenuation coefficients.
By taking the slices containing 10 or more pixels over 2,800HU CT values as metal artifact degraded images, we extract 21,625 clean slice images and 9,834 metal artifact (MA) degraded slice images from these CBCT scans.
Then we resize the extracted slice images to $400\times400$ resolutions.
Each clean CBCT slice is manually annotated with accurate tooth segmentation masks by experienced dentists.
For model training and evaluation, we employ the method in~\citep{lin2019dudonet} to simulate MA degraded images by randomly selecting the annotated segmentation masks as metal implants.
We randomly select 17,429 pairs of clean and MA degraded images from $210$ scans as the training set, 391 pairs of images from $5$ scans as the validation set, and 3,805 pairs of images from $48$ scans as the test set.

\begin{table*}[htb]
	\centering
	\begin{minipage}{1.0\linewidth}
		\centering
		\caption{\textbf{Quantitative results of PSNR (dB), SSIM~\citep{wang2004image} and Dice score our MARformer-L with different downsample ratios in the DRSA module}. ``baseline'' means the DRSA without downsampling the spatial and channel dimension. ``S$\downarrow$$r$'' or ``C$\downarrow$$r$'' indicate spatial or channel downsampling with a ratio of $r$, respectively. The FLOPs are computed on a single 400$\times$400 image as input.}
% 		\resizebox{1.0\textwidth}{!}{
		% \renewcommand\arraystretch{1.4}
        \vspace{2mm}
        \begin{tabular}{r||c|c|c|c|c|c|c|c}
        \Xhline{1.5pt}
        \rowcolor[rgb]{ .9,  .9,  .9}
        \text{Variant}  & 0  & 1 & 2 & 3 & 4 & 5 & 6 & 7\\ 
        \hline
        \text{Description} & MA & baseline & S$\downarrow$2 & C$\downarrow$2 & S$\downarrow$2~~C$\downarrow$2 & S$\downarrow$4~~C$\downarrow$4 & S$\downarrow$8~~C$\downarrow$8 & S$\downarrow$16~~C$\downarrow$16\\ \hline
        \text{PSNR} & 25.72 & 43.04  & 43.09  & 43.13  & 43.11   & 43.10 & 43.01 & 42.92\\ 
        \hline
        \text{SSIM} & 0.7207 & 0.9787  & 0.9788  & 0.9788 & 0.9789  & 0.9788 & 0.9786 & 0.9784 \\ 
        \hline
        \text{Dice} & 0.6443 & 0.8030 & 0.8019  & 0.8015  & 0.8031  & 0.8025 & 0.8031 & 0.8004 \\ 
        \hline
        \text{Params(M)} & - & 13.30 & 13.30 &  11.76 & 11.76  & 10.99 & 10.61 & 10.42 \\ 
        \hline
        \text{FLOPs(G)} & - & 82.00 & 67.17 & 71.06 & 60.25  & 54.60  & 52.63 & 51.80\\ 
        \hline
        \end{tabular}
	   % }
	\label{tab:attention}
	\end{minipage}
\end{table*}

\begin{table*}[h]
	\centering
	\begin{minipage}{1.0\linewidth}
		\centering
		\caption{\textbf{Quantitative results of PSNR (dB), SSIM, and Dice score by our MARformer-L, with varing kernel size $p$ of depth-wise convolution (a) and $\gamma$ (b) in the P2FFN module }. The FLOPs are computed on a single 400$\times$400 image as input. The comparison is made with the setting $p=7$ and $\gamma=2$.
  % The FLOPs are computed on a single 400$\times$400 image as input. It is observed that our MARformer-L obtains the best PSNR, SSIM, and Dice score results with kernel size $p=7$. Besides, our MARformer-L obtains higher PSNR and SSIM results as $\gamma$ inceases, but sufferring from more parameters and computation FLOPs. We set $\gamma=2$ to balance the model complexity and performance.
		}
        \vspace{-2mm}
	    \begin{minipage}[t]{0.48\textwidth}
	    \subtable[Study of kernel size $p$.]{
		\resizebox{1\linewidth}{!}{
	        \begin{tabular}{c|c|ccc|cc}
				\Xhline{1.5pt}
				\rowcolor[rgb]{ .9,  .9,  .9}
				\text{Variant} &
				\text{\ $p$\ }
				% & \multirow{-2}*{+PFSA}
				% & \multirow{-2}*{+PFCA}
				& \text{PSNR}
				& \text{SSIM}
				& \text{Dice}
				& Params(M) & Flops(G)
				\\
				\hline
				\hline
				1 & $3$
				& 42.72 & 0.9777 & 0.8030
				& \textbf{11.46} & \textbf{55.95}
				\\
				\hline
				% \rowcolor[rgb]{ .95,  .95,  .95}
				2 & $5$
				& 42.99 & 0.9784 & 0.8020
				& 11.52 & 57.67
				\\
				\hline
				3 & $7$
				% 57
				& \textbf{43.11} & \textbf{0.9789} & \textbf{0.8031}
				& 11.76 & 60.25
				\\
				\hline
				% \rowcolor[rgb]{ .95,  .95,  .95}
				4 & $9$
				% 55
				& 43.01 & 0.9788 & 0.8027
				& 12.09 & 63.69
				\\
				\hline
			\end{tabular}
		}
		}
	    \end{minipage}
        \begin{minipage}[t]{0.48\textwidth}
        \subtable[Study of expansion factor $\gamma$.]{
		\resizebox{1\linewidth}{!}{
			\begin{tabular}{c|c|ccc|cc}
				\Xhline{1.5pt}
				\rowcolor[rgb]{ .9,  .9,  .9}
				\text{Variant} &
				\text{\ $\gamma$\ }
				% & \multirow{-2}*{+PFSA}
				% & \multirow{-2}*{+PFCA}
				& \text{PSNR}
				& \text{SSIM}
				& \text{Dice}
				& Params(M) & Flops(G)
				\\
				\hline
				\hline
				% \rowcolor[rgb]{ .9,  .9,  .9}
				% \rowcolor[rgb]{ .95,  .95,  .95}
				5 & $1$
				& 42.81 & 0.9779 & 0.8008
				& \textbf{8.48} & \textbf{41.39}
				\\
				% \rowcolor[rgb]{ .95,  .95,  .95}
				\hline
				6 & $2$
				& 43.11 & 0.9789 & \textbf{0.8031}
				& 11.76 & 60.25
				\\
				\hline
				% \rowcolor[rgb]{ .95,  .95,  .95}
				7 & $3$
				& 43.20 & 0.9792 & 0.8016
				& 15.04 & 79.10
				\\
				\hline
				% \rowcolor[rgb]{ .95,  .95,  .95}
				8 & $4$
				& \textbf{43.28} & \textbf{0.9795} & 0.8027
				& 18.32 & 97.96
				\\
				\hline
			\end{tabular}
        }
        }
        \end{minipage}
% 		\resizebox{0.48\linewidth}{!}{
%         }
	   % \begin{minipage}{0.49\linewidth}
	   % \end{minipage}
% 		\resizebox{1.0\textwidth}{!}{

	   % }
	\label{tab:ffn}
	\end{minipage}
\end{table*}

\subsection{Training and Evaluation Details}
\noindent
\textbf{Training}.
We train the three MARformers (L/B/T) for 300 epochs with a batch size of 8, by Adam optimizer with $\beta_1=0.9$ and $\beta_2=0.99$, under an $\ell_1$ loss function. The learning rate periodically starts from $10^{-3}$ and decayed to $10^{-7}$ in 30 epochs via cosine annealing decay scheme~\citep{loshchilov2016sgdr}. We perform all experiments using the Pytorch framework~\citep{paszke2019pytorch} on two NVIDIA Tesla V100 32GB GPUs.

\noindent
\textbf{Evaluation}. We calculate the average Peak Signal Noise Ratio (PSNR) and Structural Similarity (SSIM)~\citep{wang2004image} to evaluate the quantitative MAR results of comparison methods on the MA degraded images in the test set.
We also employ the Poolformer~\citep{yu2022metaformer}, trained on the 17,429 clean slices with annotated masks~\citep{yu2022metaformer}, to perform tooth semantic segmentation on the MAR images restored by our MARformer-L. We use the Dice score to evalute the segmentation results.
%  the batch size 64 for 40,000 iterations.

%----------------------
\subsection{Comparison Results}

\noindent
\textbf{Comparison methods}.
To evaluate the effectiveness and efficiency of the proposed MARformers, we compare four representative MAR methods, \ie, LI~\citep{kalender1987reduction}, NMAR~\citep{meyer2010normalized}, RDN-CT~\citep{lin2019dudonet}, CNNMAR~\citep{zhang2018convolutional}, and two image restoration Transformers, \ie, Restormer~\citep{Zamir2021Restormer} and Uformer~\citep{wang2022uformer}. The three models of Uformer, \ie, Uformer-B/S/T are all used here. The window size of the Uformer models is set to 5 to fit the input image size. We also evaluate the popular UNet~\citep{ronneberger2015u} and ResNet~\citep{ledig2017photo} with the output head revised to suit for the MAR task, denoted as UNet-MAR and ResNet-MAR, respectively. Here, UNet-MAR has 4 levels of encoder-decoders with 1, 2, 3, 4 convolution blocks (each of which contains a 3$\times$3 conv, a BN and a ReLU), respectively. The ResNet-MAR is consisted of 8 residual blocks~\citep{ledig2017photo} with feature channel fixed as 64. These modifications ensure that UNet and RDN-CT have comparable parameters with the MARformer-L and MARformer-T, respectively. All comparison methods are retrained on our training set to achieve their best performance on the validation set.

\noindent
\textbf{Results on synthetic MAR}.
In~\cref{tab:benchmark_pairedma}, we provide the quantitative results. One can see that our MARformer-L outperforms the other methods in terms of PSNR and SSIM, but needs only 11.76M parameters and 60.25G FLOPs. Note that the second best method Uformer-B has 50.42M parameters and 205.82G FLOPs. Besides, our MARformer-T achieves similar PSNR and SSIM results with Uformer-T, but needs only 0.40M parameters and 12.82G FLOPs compared to 5.24M and 25.39G for Uformer-T. Our MARformers also achieves faster inference speeds than the Uformers, though with inferior Dice scores, respectively.
The qualitative results of visual quality are presented in~\cref{visual-group1}. We observe that our MARformer-L well recovers the teeth shapes and obtains higher PSNR and SSIM results than the other comparison methods. The light-weight MARformer-L achieves similar results to Uformer-T.
All these results validate that our MARformer is more efficient than the comparison methods on dental CBCT MAR.

%% Clinical Test
\noindent
\textbf{Results on real-world MAR}.
We also compare these methods on dental CBCT images with real-world metal artifacts. The visual results on one sample are shown in~\cref{visual-clinical}.
We observe that our MARformer-T and MARformer-L well seperate the adjacent teeth and retain their shapes. This shows that our MARformers, though trained on synthetic data, are effective on real-world MAR.

%----------------------
\subsection{Ablation Study}
%% Ablation study

To further analyze the effect of DRSA and P2FFN module, we analyze the main variant of the proposed DRSA and P2FFN modules in our MARformer-L for the MAR of dental CBCT image. The main variant 

\noindent
\textbf{Dowsample ratio in our DRSA}. To study the role of spatial and channel downsampling in our DRSA, we develop a baseline of our MARformer without downsampling the spatial or channel dimension in DRSA, and variants with other dowsample ratios of $r=4,8,16$. The comparison results are shown in~\cref{tab:attention}. From the variants 1 and 3, our MARformer-L needs less parameters by reducing the channel dimension. By dowmsampling both the spatial and channel dimensions at a ratio of 2, our MARformer-L achieves slightly better performance with less parameter amount and computational costs (FLOPs). By further increasing the downsample ratio $r=4,8,16$, the performance of our MARformer-L suffers from slightly performance drop in terms of PSNR and SSIM. This demonstrates that our MARformer-L requires less parameters and computational costs (FLOPs), while being robust to the spatial or channel dimension downsampling.

% More studies on the \textbf{kernel size of the depth-wise convolution} and the \textbf{expansion factor $\gamma$} in our P2FFN are provided in the \textsl{Supplementary File}.
\noindent
\textbf{Kernel size of the depth-wise convolution and the expansion factor in P2FFN}. To study the varying factor of P2FFN, we study the varying kernel size $p$ of the depth-wise convolution and the expansion ratio $\gamma$. Results are listed in~\cref{tab:ffn}. (a) The increase of kernel size first positively affects the restoring effectiveness, but then exihibits a subsequent decline effect. (b) While the rise of $\gamma$ holds considerable significance, it concurrently leads to a substantial growth in both the number of parameters and the computational complexity. 
It is observed that our MARformer-L obtains the best PSNR, SSIM, and Dice score results with kernel size $p=7$. Besides, our MARformer-L obtains higher PSNR and SSIM results as $\gamma$ inceases, but sufferring from more parameters and computation FLOPs. We set $\gamma=2$ to balance the model complexity and performance.

\section{Conclusion}

In this paper, we proposed an efficient MARformer for metal artifacts reduction on dental CBCT images. This is achieved by designing a new Dimension-Reduced Self-Attention (DRSA) module to reduce the computational complexity of the self-attention in Transformers. Besides, we also developed a useful Patch-wise Perceptive Feed Forward Network (P2FFN) to perceive local image information for detail reconstruction. Experiments on a large-scale dataset containing CBCT images with synthetic and real-world metal artifacts demonstrate that, our MARformer outperforms the state-of-the-art MAR methods and two image restoration Transformers on both objective metrics and visual quality.

\section*{Acknowledgments}
This work was partially supported in part by National Natural Science Foundation of China (No. 62002176 and  62176068), and the Open Research Fund (No. B10120210117-OF03) from the Guangdong Provincial Key Laboratory of Big Data Computing, The Chinese University of Hong Kong, Shenzhen. 
\normalsize
\bibliography{references}

%%%%%%%%%%%%  Supplementary Figures  %%%%%%%%%%%%
%\clearpage

%%%%%%%%%%%%%%%%   End   %%%%%%%%%%%%%%%%
%\end{multicols}  % Method B for two-column formatting (doesn't play well with line numbers), comment out if using method A
\end{document}